\documentclass[prb,showpacs]{revtex4}
\usepackage{graphicx}

\newcommand{\ve}{\varepsilon}
\newcommand{\p}{\prime}
\newcommand{\wf}{\omega_f}
\begin{document}
\title{Relaxation between electrons and surface phonons in nanoscale metal films} 
\author{Navinder Singh}
\email{navinder@iopb.res.in}
\affiliation{Institute of Physics, Bhubaneswar 751005, India}
\begin{abstract}
The Two Temperature model of M I Kaganov, I M Lifshitz and L V Tanatarov (Sov. Phys. JETP 4, 173 (1957)) about relaxation between electrons and bulk phonons is extended to the case of surface phonons. The state of electrons and phonons is described by equilibrium Fermi and Bose functions with different temperatures. 
The new feature added is the geometric constraint of surface phonons. We obtain expressions for the energy transfer rate from degenerate hot electrons to surface-phonons, which is order of magnitude less than that for the bulk.
\end{abstract}
\maketitle
\noindent
{\bf INTRODUCTION}\\
In nanoscale metallic systems such as island metal films used in micro-electronics[1], the phenomenon of hot electron scattering by surface phonons is quite important. One important problem in this field is to calculate the energy transfer rate between excited hot electrons and the lattice bath. The present paper (an extention of the work of M I Kaganov, I M Lifshitz and L V Tanatarov [2]) is devoted to the calculation of energy transfer rate from degenerate hot electrons to surface phonons. The whole process occurs at pico-second time scales[3]. We consider the case of a homogeneously photoexcited nano-scale metal film. The film thickness is of the order of electron mean free path. So, the electron surface-phonon interaction is important. We also consider the strong damping of surface phonons, due to the strong coupling with the substrate on which film was developed. Thus, there is no surface standing modes at the film surface.
\vspace{0.40cm}

\noindent
{\bf THE MODEL}\\
Consider a hot Fermi electron distribution at temperature $T_e$ and a phonon(2-D) equilibrium distribution at temperature $T \;(T<T_e)$. Both degenerate distibutions are weaky interacting subsystems. The energy flows from hot degenerate electron distribution to phonon bath. In the following, we calculate, in line with Kaganov[2],the energy transfered per second per unit volume from the hot degenerate electron distribution to the relatively cold phonon gas. The equilibrium distributions for electrons and phonons are
\begin{eqnarray}
&&N_k = \frac{1}{e^{\beta_e(\ve-\ve_0)} +1} \beta_e =\frac{1}{k_B T_e}
\nonumber\\
&&N_f = \frac{1}{e^{\beta\hbar\omega_f} - 1}\beta =\frac{1}{k_B T} .
\end{eqnarray}
\begin{figure}
\includegraphics[height = 6 cm,width = 14 cm]{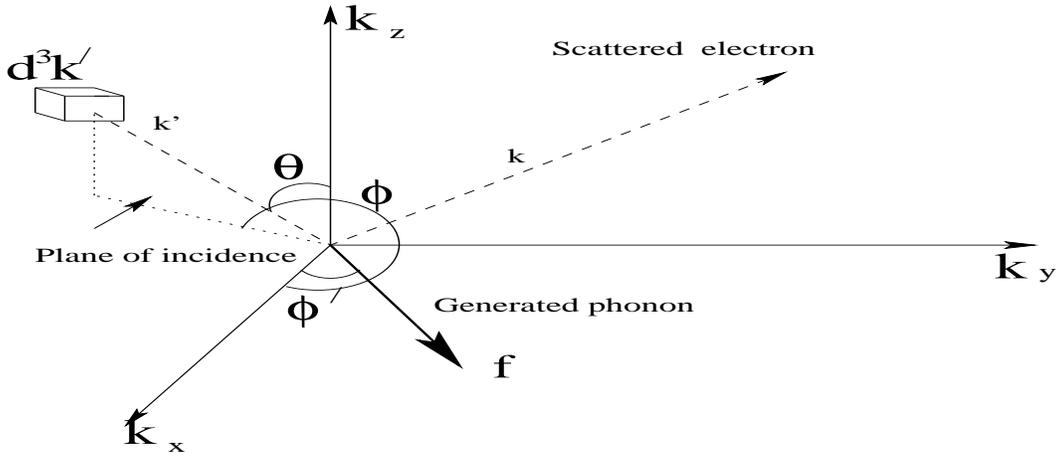}
\caption{Electrons scattering from the metal surface.}
\end{figure}
Energy and momentum conservation conditions gives(Figure 1)
\begin{equation}
\begin{array}{cc}
\ve_{k^\p} - \ve_{k} = \hbar\omega \;\;,{\bf k^\p- k = f}\;\;,  k^\p_x -k_x = f_x \;,\\
\; k^\p_y - k_y = f_y \;\,,\,k^\prime_z=-k_z\;\;,\omega = s f \;,\\
\ve_{k^\p}=\frac{\hbar^2{k^\p}^2}{2m}\;,\;\frac{\hbar^2}{2m}(2 [{k^\p}_x f_x + {k^\p}_y f_y]- f^2)=\hbar sf\;, \\
\frac{\hbar^2}{2m}[{k^\prime_x}^2-(k^\prime_x-f_x)^2 + {k^\prime_y}^2 - 
(k^\prime_y-f_y)^2] = {\hbar}sf\,.
\end{array}
\end{equation}
On simplifying
\begin{equation}
\frac{\hbar^2}{2m}[2k^{\prime}\sin{\theta}\cos(\phi-\phi^{\prime})-f] = {\hbar}s.
\end{equation}
Where ${\phi}$ is the angle between $k_x$-axis and plane of incidence. ${\phi^{\p}}$ are the angle 
between scattered phonon direction and $k_x$-axis, and ${\theta}$ between incident electron 
direction and $k_z$  direction as shown in Figure 1. The probability $W$ per unit time that the electron in a state with wave vector $ k^\prime $
will scatter to a state with wave vector $k$ by  emitting a phonon of wave vector $f$ is;
\[W(k^\prime-f;k^\prime)=\alpha\delta(\ve_{k^\p} - \ve_k - \hbar
\omega),\,\,\alpha=(\pi U_s^2/\rho V S_s^2).\]
With $U_s$ as the electron surface-phonon interaction constant. Here ${\rho}$, V and $S_s$ is the metal density, unit cell volume and surface sound speed respectively. The change per unit time per unit volume in the number of surface-phonons with wave vector $f$ and energy $\hbar\omega$ is (Bloch-Boltzmann-Peierls formula);

\begin{equation}
\dot{N}_f = \int\alpha\omega_f\{(N_{f} +1)N_{k^{\prime}}(1-N_{k})-
N_fN_k(1-N_{k^{\prime}})\}\delta(\ve_{k^{\prime}}-\ve_{k}-\hbar\omega)
(2/(2\pi)^3)d\tau_{k^{\prime}}.
\end{equation}
Using the energy and momentum conservation equations, the delta function can be written as
$\delta (\ve_{k^\prime} -\ve_k - \hbar\omega)=\frac{2 m}{\hbar^2 f}\delta 
[(2 k^\prime \sin \theta \cos (\phi-\phi^\prime)-f)-\frac{2 m s}{\hbar}]$, and for a metal, 
we have $(f\sim 10^9  m^{-1})\gg (\frac{2 m s}{\hbar}\sim 10^7  m^{-1}) $. So the above 
equation i.e.,(Eq.4) is
\begin{eqnarray}
&&\dot{N}_f = \int\alpha\omega_{f}\{(N_{f} +1)N_{k^{\prime}}(1-N_{k})-
N_fN_k(1-N_{k^{\prime}})\}\delta(2k^{\prime}\sin{\theta}\cos(\phi-\phi^{\prime})-f)\nonumber\\&& 
\times\frac{4m{k^\prime}^2}{2\pi^2\hbar^2 f}\sin{\theta}d{\theta}d{\phi}dk^{\prime}.
\end{eqnarray}
\begin{eqnarray}
&&\dot{N}_f = \left[\frac{4 m\alpha s \hbar}{(2\pi\hbar)^3}\right] \int_{k_m}^
{\infty}[(N_f +1)N_{k^\prime}(1-N_k)- N_f N_k (1-N_{k^\prime})]{k^\prime}^2 dk^{\prime}
\nonumber\\ 
&& \times \int_0^{\pi/2}\sin \theta d\theta\int_0^{2\pi}\delta[2k^\prime\sin
\theta\cos(\phi-\phi^\prime)-f]d\phi.
\label{6.6}
\end{eqnarray}
The last integral in the above equation is
\begin{eqnarray}
&&\int_0^{2\pi}\delta[2k^\prime\sin\theta\cos(\phi-\phi^\prime)-f]d\phi=\frac{1}{|2k^
{\prime}\sin\theta|}\nonumber\\ &&\times\left\{\frac{1}{|\sin(\phi_1-\phi^\prime)|}\int_
0^{2\pi}\delta(\phi-\phi_1)d\phi + \frac{1}{|\sin(\phi_2-\phi^\prime)|}\int_0^{2\pi}\delta
(\phi_2-\phi^\prime)d\phi \right\}\nonumber\\&& = \frac{1}{k^\prime\sin\theta\sqrt(1-
f^2/(4{k^\prime}^2\sin^2\theta))}
\label{6.7}{\rm ~inserting~this~in~(6)~we~get}
\end{eqnarray}
\begin{equation}
\dot{N}_f = \left[\frac{ m\alpha s }{(2\pi\hbar)^2}\right] \int_{k_m}^
{\infty}[(N_f +1)N_{k^\prime}(1-N_k)- N_f N_k (1-N_{k^\prime})]{k^\prime} dk^{\prime}
\label{6.8}
\end{equation}
The above mentioned process will always happen as from energy and momentum conservation,
$\sin{\theta} \simeq f/2k^{\prime}$, which holds good in a metal as $f<k^\prime$.
By inserting for $N_e$ and $N_f$ in equation (8) we get
\begin{equation}
\dot{N}_f = \left[\frac{ m\alpha s }{(2\pi\hbar)^2}\right]\left(\frac{e^{\beta\hbar\wf}-
e^{\beta_e\hbar\wf}}{e^{\beta\hbar\wf} - 1}\right)\int_{k_m}^\infty\frac{e^{\beta_e(\ve_k^\p
 -\hbar\wf-\ve_0)}k^\p dk^\p}{(e^{\beta_e(\ve_k^\p-\ve_0)}+1)(e^{\beta_e(\ve_k^\p-\hbar\wf-
\ve_0)}+1)}
\end{equation}
Here, we will make an approximation to solve the integral in the above equation. The first 
approximation is that the phonon energy $\hbar\omega_f(meV) \ll k_B T_e(eV)$, the electron energy.
So, $\beta_e\hbar\wf\sim 0 $. With this the integral in Eq.(9) is
\[\frac{m}{\beta_e\hbar^2}\left[\frac{1}{e^{f(k_m)} +1}\right]\;\;,\;f(k_m) = 
\beta_e\left[\frac{\hbar^2 k_m^2}{2m}-\ve_0\right].\]
As $|\frac{\beta_e\hbar^2 k_m^2}{2m}|\ll|\beta\ve_0|$, the quantity in the square brackets is 
order of unity. Finally, the integral in Eq.(9) is $\frac{m\wf}{\hbar}(\frac{1}{\beta_e\hbar
\wf})\sim \frac{m\wf}{\hbar}(1/(e^{\beta_e\hbar\wf}-1)) $. With all this Eq.(9) takes the form
\begin{equation}
\dot{N}_f = \left[\frac{ m^2\alpha s \wf}{(2\pi\hbar)^2\hbar}\right]\left(\frac{e^{\beta\hbar\wf}-
e^{\beta_e\hbar\wf}}{(e^{\beta\hbar\wf}-1)(e^{\beta_e\hbar\wf}-1)}\right).
\label{6.10}
\end{equation}
The energy transfered by the electrons to the surface-phonons per unit volume per unit time is
\begin{equation}
U_{surface} = \frac{a^2}{(2\pi)^2}\int_0^{f_{Ds}}\dot{N}_f\hbar\wf 2\pi f df,\;\;a = lattice\;\;constant,
\end{equation}
where $f_{Ds}$ is  the Debye wave vector for the surface phonons. From Eq.(10) and Eq.(11) with relations $\omega_{Ds} = S_s f_{Ds} \;\;,\;\hbar \omega_{Ds} = k_B T_{Ds}$ and setting $x = \hbar\wf/k_BT_e$, we get
\begin{eqnarray}
&&U_{surface} = \left[\frac{\pi U_s^2 m^2}{(2\pi)^3\hbar^2\rho a S_s^3}\right]\left(\frac{k_BT_{Ds}}{\hbar}\right)^4\nonumber\\
&&\times\left[\left(\frac{T_e}{T_{Ds}}\right)^4\int_0^{T_{Ds}/T_e}\frac{x^3}{e^x-1}dx-
\left(\frac{T}{T_{Ds}}\right)^4\int_0^{T_{Ds}/T}\frac{x^3}{e^x-1}dx \right].
\end{eqnarray}
Here, $T_{Ds}$ is the surface Debye temperature. Now the equation (12) can be simplified in two special cases, first, for low electron and phonon temperatures as compared to Debye temperature, i.e., $T, T_e \ll T_{Ds}$ , Eq.(12) reduce to
\begin{equation}
U_{surface} = \left[\frac{\pi U_s^2 m^2}{(2\pi)^3\hbar^2\rho a S_s^3}\right]\left(\frac{k_BT_{Ds}}{\hbar}\right)^4\left[\frac{T_e^4 - T^4}{T_{Ds}^4}\right]\int_0^\infty\frac{x^3}{e^x-1}dx.
\end{equation}
{\bf An important point to be noted in the above equation is that the electron to phonon 
energy transfer rate depends upon $4^{th}$ power of electron and phonon temperatures as 
compared to the corresponding case in the bulk(there it is $5^{th}$ power of electron and 
phonon temperatures[2])}. In second special case, when $T_e\;,\;T\gg T_{Ds}$ , we get
\begin{equation}
U_{surface} = \left[\frac{\pi U_s^2 m^2}{3(2\pi)^3\hbar^2\rho a S_s^3}\right]\left(\frac{k_BT_{Ds}}{\hbar}\right)^4\left[\frac{T_e - T}{T_{Ds}}\right].
\label{6.14}
\end{equation}
The above equation (Eq.(14)) is the basics of what is called the two temperature model. The surface Debye temperature $T_{Ds} = \frac{h}{k_B} f_{Ds}$, for two acoustic modes per atom is given by
\begin{equation}
\frac{L^2}{(2\pi)^2}\int_0^{f_{Ds}} 2\pi fdf = 2 N_{surface},
\end{equation}
which gives $f_{Ds} = \sqrt{8\pi n^{2/3}}$ , $n$ is the number density per unit volume. Now, for the bulk case [2]
\begin{equation}
U_{bulk} = \left[\frac{m^2 U_b^2\omega_{Db}^4k_B}{2(2\pi)^3\hbar^3\rho S_b^4}\right][T_e - T].
\end{equation}
From  Eq.(14) and Eq.(16) we have
\begin{equation}
\frac{U_{surface}}{U_{bulk}} = \frac{2}{3}\pi(8\pi)^{3/2}\left[\frac{U_s}{U_b}\right]^2\frac{n S_b^4}{a \omega_{Db}^4}.
\label{6.17}
\end{equation}
For a gold metal film, assuming $U_s = U_b$, with $a = 4.1\times10^{-10} m\;,\;\rho = 19.3
\times10^{3} kg/m^3\;,\;n = 5.9\times10^{28} m^{-3}\;,\;T_D = 185\;K\;,\;\omega_D = 2.42
\times10^{13} rads/sec$, the above ratio is $\sim 0.088$ or about 9 percent.
\vspace{0.40cm}

\noindent
{\bf DISCUSSION}\\
We have treated here the problem of energy relaxation of photo excited degenerate electrons in a nano-scale metal film. The new feature added is the geometric constraint of surface phonons which is responsible for reduced energy transfer rate in nano-scale metal films as compared to bulk metals. This is a simple calculation. The effect of electron-phonon screening and quantum confinement effects due to nano size, are not taken into account.
\vspace{0.40cm}

\noindent
{\bf ACKNOWLEDGMENT}\\

The author would like to thank Prof. R. Srinivasan for technical support and discussion.

\end{document}